\documentclass[iop]{emulateapj}

\newcommand{\rxte}{{\it RXTE}}

\newcommand{\rxtepca}{{\it RXTE}/PCA}
\newcommand{\rxtehexte}{{\it RXTE}/HEXTE}
\newcommand{\swift}{{\it Swift}}

\newcommand{\swiftxrt}{{\it Swift}/XRT}

\newcommand{\heasoft}{{\sc heasoft}}

\newcommand{\isis}{{ISIS}}

\newcommand{\diskbb}{{\sc diskbb}}
\newcommand{\phabs}{{\sc phabs}}
\newcommand{\cutoffpl}{{\sc cutoffpl}}
\newcommand{\powerlaw}{{\sc powerlaw}}
\newcommand{\gaussian}{{\sc gaussian}}
\newcommand{\edge}{{\sc edge}}

\shorttitle{BH plasma conditions and outflow properties}
\shortauthors{Koljonen et al.}

\begin{document}


\title{A connection between plasma conditions near black hole event horizons and outflow properties}


\author{K. I. I. Koljonen\altaffilmark{1}, D. M. Russell\altaffilmark{1}, J. A. Fern\'andez Ontiveros\altaffilmark{2}, Sera Markoff\altaffilmark{3}, T. D. Russell\altaffilmark{4}, J. C. A. Miller-Jones\altaffilmark{4}, A. J. van der Horst\altaffilmark{5}, F. Bernardini\altaffilmark{1}, P. Casella\altaffilmark{6}, P. A. Curran\altaffilmark{4}, P. Gandhi\altaffilmark{7}, R. Soria\altaffilmark{4}}
\affil{$^{1}$New York University Abu Dhabi, PO Box 129188, Abu Dhabi, UAE}
\affil{$^{2}$Istituto di Astrofisica e Planetologia Spaziali (INAF--IAPS), Via Fosso del Cavaliere 100, 00133 Roma, Italy}
\affil{$^{3}$Astronomical Institute `Anton Pannekoek', University of Amsterdam, PO Box 94249, 1090 GE Amsterdam, Netherlands}
\affil{$^{4}$International Centre for Radio Astronomy Research -- Curtin University, GPO Box U1987, Perth, WA 6845, Australia}
\affil{$^{5}$Department of Physics, The George Washington University, 725 21st Street NW, Washington, DC 20052, USA}
\affil{$^{6}$INAF, Osservatorio Astronomico di Roma, Via Frascati 33, I-00040 Monteporzio Catone, Italy}
\affil{$^{7}$School of Physics \& Astronomy, University of Southampton, Highfield, Southampton SO17 1BJ, UK}


\altaffiltext{1}{email: karri.koljonen@nyu.edu}


\begin{abstract}

Accreting black holes are responsible for producing the fastest, most powerful outflows of matter in the Universe. The formation process of powerful jets close to black holes is poorly understood, and the conditions leading to jet formation are currently hotly debated. In this paper, we report an unambiguous empirical correlation between the properties of the plasma close to the black hole and the particle acceleration properties within jets launched from the central regions of accreting stellar-mass and supermassive black holes. In these sources the emission of the plasma near the black hole is characterized by a power law at X-ray energies during times when the jets are produced. We find that the photon index of this power law, which gives information on the underlying particle distribution, correlates with the characteristic break frequency in the jet spectrum, which is dependent on magnetohydrodynamical processes in the outflow. The observed range in break frequencies varies by five orders of magnitude, in sources that span nine orders of magnitude in black hole mass, revealing a similarity of jet properties over a large range of black hole masses powering these jets. This correlation demonstrates that the internal properties of the jet rely most critically on the conditions of the plasma close to the black hole, rather than other parameters such as the black hole mass or spin, and will provide a benchmark that should be reproduced by the jet formation models.

\end{abstract}



\section{Introduction}

Powerful jets of plasma are produced by accreting black holes of all sizes ranging from stellar-mass black holes in Galactic X-ray binaries (XRBs) to supermassive black holes in active galactic nuclei (AGN). The jets are launched close to the black hole event horizon, but the conditions leading to jet formation are still debated. Several models have been put forward that predict the jet properties are governed by the accretion rate, black hole spin, the magnetic field strength and configuration, and/or the properties and location of the inner accretion flow \citep{blanzn77,blanpa82,meie01,tcheet11}. Despite the recent advances in our understanding of jets, the total jet power is notoriously hard to measure because rather than being radiated locally, the bulk of the energy is transported to large distances from the black holes as a dark flow. Nevertheless, there are estimates that imply jets can dominate the power output of black holes \citep{gallet05,ghiset14}, and that a large fraction of the mass in the accretion flow can escape via outflows depending on the state of accretion \citep{neille09,pontet12}.

The classic signature of a relativistic, compact jet is a flat or slightly inverted ($\alpha_{thick} \geq 0$ where $S_{\nu} \propto \nu^{\alpha_{thick}}$) radio spectrum composed of overlapping synchrotron spectra from different locations in the jet \citep{blanko79}. The flat/inverted radio spectrum will break at some higher frequency $\nu_{b}$ associated with a transition to low optical depth, either at the base of the jet (e.g. \citealt{blanko79,konigl81,ghiset85}) or at the location where particles are accelerated due to the presence of a shocked zone (e.g. \citealt{market05,marschet08,polket14}). The slope of the optically thin spectrum is usually close to $\alpha_{thin} \sim -0.7$ or steeper if the electrons are cooled or have a thermal distribution of energies \citep{peer09}. The radio luminosity is often used as a proxy for jet power, but jet powers inferred this way can disagree with estimates derived from jet feedback in the form of large-scale lobes and cavities, by orders of magnitude \citep{kordet08}. While the jet power is often estimated directly from the radio luminosity with either a linear or power law dependency, the total luminosity of the jet is dominated by the high frequency emission and can only be measured accurately by observing the whole spectrum of the jet.

Up until recently, the reason why only a few jet breaks had been identified from XRBs, was due to the companion star or accretion disc dominating the emission around the break frequencies \citep{gallo07,migliari07,rahoui11}, and the lack of mid-IR data acquired from these sources. However, the latter has recently been improved with multiwavelength campaigns of black hole outbursts including mid-IR and mm data. A recent key discovery shows that the peak flux density of the jet can vary dramatically with the state of accretion of XRBs, while the radio luminosity can remain steady \citep{russet13b,vandet13,russet14}, casting into doubt the reliability of using the radio luminosity as a proxy for the jet power. Similarly for most AGN, the break and most of the jet spectrum lie under other non-synchrotron components, e.g. galaxy, accretion disc and/or torus, and only recently owing to the use of adaptive optics at optical/IR frequencies a core jet spectrum has been revealed from a few close-by low-luminosity AGN \citep{fernandez12}.

Continuously-launched, flat-spectrum jets are commonly observed during hard and intermediate X-ray states of XRBs \citep{fendga14}, when the X-ray spectrum is dominated by a power law spectral component. This component is generally thought to represent the inverse Comptonization of soft seed photons in a plasma cloud of hot electrons close to the black hole. These hot electrons are thought to be located near the black hole, whether in a thermal Comptonization dominated accretion flow \citep{zdziarski98}, radiatively inefficient accretion flow (ADAF/RIAF; \citealt{yuanet03}) or at the base of the jet \citep{market05}. Thus, they are usually dubbed as the corona, a term which embodies all the different possibilities for the origin of the hot electrons. Due to the simple nature of black holes, accretion physics is expected to scale globally with black hole mass. Similar to XRBs, the hard X-ray emission from AGN is expected to arise from a Comptonizing corona \citep{haardt93}, and evidence for its compact size and location close to the central black hole has come from several research avenues including studies of iron line spectra and variability (e.g. \citealt{fabian09}), reverberation (e.g. \citealt{uttley14}), microlensing (e.g. \citealt{morgan12}), and obscuration of the corona by clouds (e.g. \citealt{sanfrutos13}).    

As noted above, the existence and power of jets launched via accretion onto stellar-mass black holes in XRBs has been found to be linked with specific X-ray spectral and timing properties, which trace the nature of mass accretion onto the black hole. Thus we can expect a link between accretion and ejection to be present in their constituent components. In this paper we test this link using broadband spectral energy distributions of stellar-mass and supermassive black holes. The structure of the paper is the following. In Section 2, we present in detail the multiwavelength properties of our sample of XRBs and AGN, and in particular the spectral energy distributions (SEDs) from the AGN sample, and the reduction and analysis of the X-ray data from XRBs. The results of the analysis, which show an anti-correlation between the X-ray power law photon index and the jet break frequency are presented in Section 3. In Section 4, we discuss the origin of this correlation and whether or not the system parameters have an effect to it. In Section 5, we present our conclusions and discuss the ramifications of our results.   

\section{Observations and data analysis}

We searched for sources that have a clear X-ray view to the central region of the black holes and a well-sampled jet spectrum, in systems where the measurements are not likely to be skewed by relativistic beaming. This search resulted in eleven stellar-mass black holes in XRBs, and seven low accretion rate AGN that all show an unambiguous, isolated jet spectrum in addition to a well-defined X-ray spectrum that is quasi-simultaneous for XRBs. All sources in our sample exhibit the classic signature of a relativistic, flat or slightly inverted compact jet. For each source we fit a broken power law to determine the frequency of the jet break $\nu_{b}$ and the flux density $S_{\nu,b}$ at the break frequency based on a broad range of multiwavelength observations.

\subsection{XRBs}

\subsubsection{Radio and jet break data}

We searched the literature for jet break frequency values from the SEDs of black hole XRBs that have simultaneous X-ray observations. The search resulted in nine sources in the hard X-ray state from \citet{russet13a} and references therein: 4U 1543--47, Cyg X--1, GS 1354--64, GX 339--4, V404 Cyg, V4641 Sgr, XTE J1118$+$480, XTE J1550--564 and XTE J1752--223, with two additional intermediate X-ray state sources MAXI J1659--152 \citep{vandet13} and MAXI J1836--194 \citep{russet14}, in which flat spectrum radio jets were still observed but the X-ray spectra were softer (see the referenced papers for the figures of the SEDs). The values for the jet break frequencies, radio and jet break fluxes from the XRB sample are tabulated in Table 1. In some cases the location of the jet break could not be well constrained. We therefore used multiwavelength data to restrict the jet break to lie in a certain frequency range in order to include them in the Monte Carlo bootstrap estimation of the correlation and linear regression (Section 3). For GS 1354--64 and V4641 Sgr the optically thick radio-to-IR spectrum overshoots the X-ray spectrum, and thus the jet spectrum has to break before the X-ray regime. For GX 339--4 and MAXI J1659--152 the optically thick radio spectrum overshoots the data points in the IR, and thus the jet spectrum has to break before the IR regime. For XTE J1752--223 the optically thin jet spectrum overshoots the upper limit in the radio, and thus the jet break has to be between the radio and IR regimes. We use the most conservative ranges possible based on the existing data, i.e. we do not assume any particular slope for the optically thin spectrum. In the case of XTE J1752--233 we assume that the spectral index of an optically thick, self-absorbed spectrum can not be more than 5/2 based on the synchrotron theory.     

\begin{table*} \centering
\caption{Literature values for the jet break frequencies, radio and jet break fluxes in XRBs. The columns are: (1) source name, (2) starting time of the observation in MJD, (3) logarithm of the jet break frequency, (4) flux density at the jet break, (5) radio flux density at 5 GHz, (6) the excess luminosity over the radio luminosity up to the jet break $L_{\rm b}/L_{\rm 5GHz} = \nu_{b} S_{\nu,b}/\nu_{\rm 5GHz} S_{\rm \nu,5GHz}$, and (7) references for the jet break values (R13a: \citet{russet13a}, Ra11: \citet{rahoui11}, G11: \citet{gandet11}, C13: \citet{corbet13}, vdH13: \citet{vandet13}, R14: \citet{russet14}, C11: \citet{chaty11}, R12: \citet{russell12}).} 
\label{jetbreaks}
\begin{tabular}{llccccc} \tableline
Source & Time & log $\nu_{b}$ & S$_{\nu,b}$ & S$_{\nu,5 GHz}$ & log $L_{\rm b}/L_{\rm 5GHz}$ & Ref \\
& MJD & Hz & mJy & mJy \\ \tableline
4U 1543--47 & 52490.00 & 13.98$\pm$0.20 & 9.2$\pm$2.8 & 4.00$\pm$0.05 & 4.64$\pm$0.34 & R13a \\
Cyg X--1 & 53513.00 & 13.45$\pm$0.02 & 16.8$\pm$1.2 & 15.6$\pm$0.2 & 3.78$\pm$0.06 & R13a,Ra11 \\
GS 1354--64 & 50772.00 & [14.13--18.00] & 2.3$\pm$0.1 & 2.8$\pm$0.1 & [4.31--8.25] & R13a \\
GX 339--4 & 50648.00 & 14.26$\pm$0.12 & 10.7$\pm$0.9 & 14$\pm$3 & 4.44$\pm$0.25 & R13a,G11 \\ 
& 55266.00 & 13.65$\pm$0.24 & 115$\pm$11 & 9.1$\pm$1.1 & 5.05$\pm$0.33 & R13a,G11 \\
& 55617.00 & [12.63--14.26] & 21.0$\pm$1.0 & 2.54$\pm$0.04 & [3.83--5.51] & C13 \\
MAXI J1659--152 & 55467.10 & [10.34--14.67] & 10$\pm$1 & 10.5$\pm$0.8 & [0.54--5.03] & vdH13 \\
& 55467.90 & [10.63--14.26] & 11.2$\pm$0.6 & 9.9$\pm$0.3 & [0.95--4.65] & vdH13 \\
& 55470.10 & 10.34$^{+0.09}_{-0.11}$ & 8.35$\pm$0.45 & 9.75$\pm$0.30 & 0.57$\pm$0.14 & vdH13 \\
& 55473.90 & [11.54--14.68] & 10.5$\pm$3.2 & 3.65$\pm$0.09 & [2.13--5.57] & vdH13 \\
& 55476.80 & 10.34$^{+0.09}_{-0.11}$ & 0.41$\pm$0.07 & 0.63$\pm$0.03 & 0.45$\pm$0.20 & vdH13 \\
& 55488.80 & 9.69$^{+0.08}_{-0.10}$ & 0.23$\pm$0.03 & 0.23$\pm$0.03 & -0.01$\pm$0.20 & vdH13 \\
MAXI J1836--194 & 55807.12 & 11.37$^{+0.11}_{-0.27}$ & 415$^{+700}_{-190}$ & 29$\pm$1 & 2.83$\pm$0.55 & R14 \\
& 55821.97 & 11.16$\pm$0.55 & 64$\pm$16 & 34.5$\pm$0.9 & 1.73$\pm$0.67 & R14 \\
& 55830.95 & 11.98$^{+0.27}_{-0.21}$ & 260$^{+140}_{-45}$ & 14.1$\pm$0.4 & 3.55$\pm$0.34 & R14 \\
& 55846.01 & 12.74$^{+0.13}_{-0.02}$ & 185$^{+30}_{-15}$ & 5.5$\pm$0.4 & 4.57$\pm$0.15 & R14 \\
& 55861.00 & 13.71$^{+0.37}_{-0.01}$ & 27$^{+18}_{-5}$ & 2.5$\pm$0.3 & 5.04$\pm$0.40 & R14 \\
V404 Cyg & 47728--9 & 14.26$\pm$0.06 & 178$\pm$16 & 17.0$\pm$0.7 & 5.58$\pm$0.12 & R13a \\
V4641 Sgr & 52857.00 & [14.67--18.00] & 93$\pm$46 & 621$\pm$2 & [3.85--7.65] & R13a \\
XTE J1118$+$480 & 51649.00 & 13.43$\pm$0.09 & 290$\pm$65 & 4.7$\pm$0.7 & 5.52$\pm$0.25 & R13a \\
& 53386.00 & 12.65$\pm$0.08 & 170$\pm$19 & 4.4$\pm$0.2 & 4.54$\pm$0.15 & R13a \\
XTE J1550--564 & 51697.00 & 13.68$\pm$0.33 & 38$\pm$27 & 0.9$\pm$0.1 & 5.61$\pm$0.76 & R13a \\
XTE J1752--223 & 55378.00 & [10.99--14.26] & 1.3$\pm$1 & $<$0.3 & $<$5.16 & R13a,R12 \\
\tableline
\end{tabular}
\end{table*}      

\subsubsection{X-ray data}

We used \rxte\/ (Rossi X-ray Timing Explorer) and \swift\/ X-ray observatories to select pointings from the High Energy Astrophysics Science Archive Research Center (HEASARC) archive which are as contemporaneous with the radio observations as possible (within a day for all sources with the exception of GS 1354--64, where the nearest pointing was found to be within two days). The only exception was V404 Cyg, which has no \rxte\/ or \swift\/ data available, and thus we collected the relevant spectral modeling values from the literature \citep{zycki99}. Each \rxte\/ pointing was individually reduced by the standard method as described in the \rxte\/ cookbook using \heasoft\/ 6.16. The Proportional Counter Array (\rxtepca) spectrum was extracted from all available proportional counter units (PCU) in each pointing to maximize the photon counts in the spectra, excluding PCU--0 and PCU--1 after their propane loss in 13 May 2000 and 25 December 2006 respectively. The High Energy X-ray Timing Experiment (\rxtehexte) spectrum was extracted from both clusters A and B when available. After 14 December 2009 when cluster B stopped rocking we used cluster A for the source data and estimated the background using cluster B. \rxtepca\/ spectra were then grouped to a minimum of 5.5 sigma significance per bin, and bins below 3.5 keV and above 20 keV were ignored. In addition, we added 0.5\% systematic error to all channels. In similar fashion \rxtehexte\/ spectra were grouped to a minimum of five sigma significance per bin (20 sigma per bin after 14 December 2009), and bins below 18 keV and above 200 keV were ignored. In addition, we added 1\% systematic error to all channels of spectra taken after 14 December 2009. For MAXI J1659--152 and MAXI J1836--194 we selected additionally simultaneous \swift\/ spectra (within a day from \rxte\/ pointings) to better gauge the effect of the disc component on the X-ray spectra. The X-Ray Telescope (\swiftxrt) windowed timing (WT) mode data was processed using {\sc xrtpipeline} in \heasoft\/ 6.16, and subsequently the source and background spectrum and response files were extracted using {\sc xrtproducts}. Exposure maps were generated for each pointing and the pile-up was taken into account by excluding a circular region at the source position, with the region size depending on the count rate \citep{reynolds13}. 

\subsection{AGN}

For the majority of AGN, both the break and most of the jet spectrum lie under other non-synchrotron components (e.g. emission from the galaxy and the star-forming regions). To isolate the true core emission we need observations based on high-angular resolution techniques: Hubble Space Telescope (HST) in the optical, ground-based adaptive optics observations in the near-IR, and ground-based diffraction limited observations in the mid-IR. Our sample of seven AGN consists of four low-luminosity AGN, one FR-I, one FR-II and Sgr A*. The four low-luminosity AGN are the brightest and nearest low-luminosity AGN (L$_{\rm bol} \lesssim 10^{42}$ erg/s) accessible from the Southern Hemisphere, and correspond to those targets with successful adaptive optics observations and the best HST coverage in the optical and UV range. These targets were extracted from the project ``The central parsecs of the nearest galaxies'' \citep{prieto10,reunanen10}, a high-spatial resolution study of the brightest and nearest AGN carried out at sub-arcsecond scales with the Very Large Telescope, using the NaCo and VISIR instruments in the near- and mid-IR ranges, respectively. Three of the targets are the canonical reference for the definition of the low-luminosity AGN class: NGC 1052 \citep{heckman80}, NGC 1097 \citep{keel83}, and M87 \citep{fabian95}. Together with the Sombrero galaxy (NGC 4594), this sample is the best representation of the low-luminosity class in the nearby universe, also in terms of host galaxy (Sa, SB(s)b, E, and S0 for NGC 1052, NGC 1097, M87, and Sombrero respectively). Furthermore, the SEDs of these four objects are in agreement with previous works \citep{elvis94,ho99,eracleous10} in the common wavelength ranges covered (X-rays, optical/UV, and radio), but the lack of high-angular resolution near- and mid-IR observations in the past prevented the identification of a jet-dominated continuum in these sources. Similarly, 3C 120 and Cygnus A are representatives of the radio galaxy population and were found to have isolated jet spectra in their high resolution data. 3C 120 belongs to the FR-I class (core-dominated radio galaxies) and Cygnus A to the FR-II class (lobe dominated), and they are accreting at a relatively low accretion rate, and show unambiguous flat/inverted synchrotron radio spectra and a spectral break. In addition to the above-mentioned AGN, we add Sgr A* to the sample, which is accreting matter in a very low accretion rate. The jet break frequency and radio measurement values are taken from the literature \citep{becket02}.

The small number of low-luminosity AGN in the sample is mainly due to the faintness of their nuclei, which hinder the use of adaptive optics in these objects. Moreover, we discarded those AGN affected by internal (torus) or external obscuration (dust lanes in the host galaxy), since it is not straightforward to identify the presence of a jet-dominated continuum in those cases, e.g. Centaurus A \citep{meisenheimer07}.

\subsubsection{Radio and jet break data} 

The data set consists of sub-arcsecond measurements ($<$0.4" apertures) from radio to ultraviolet \citep{fernandez12,canalizo03,lr14,lee08,asmus14,doi13}, in addition to low-angular resolution ($>$1'' apertures) measurements from NASA/IPAC Extragalactic Database (NED), the Wide-field Infrared Survey Explorer at IPAC (WISE) archive, the Akari Point Source Catalog and the 2MASS Point Source Catalog. All data have been corrected for Galactic extinction \citep{schlafly11}. For 3C 120 we also measured fluxes from images taken from the HST Legacy Archive. The sub-arcsecond photometry ensures the extraction of the nuclear continuum, minimizing the possible contribution of extended components, e.g. the underlying galaxy and extended dust emission. The low energy part of each SED of the sample can be fit with a broken power law representing a self-absorbed synchrotron spectrum from the jet with a flat or inverted spectrum below the break frequency, as is the case for black hole XRBs. For the fits we allowed the optically thick index to be also negative to avoid/amend the influence of optical depth effects in the determination of the break frequency and the associated flux. The spectral slope at higher frequencies than the jet break in some of the AGN is extremely steep ($\alpha_{thin}\sim-4$), which could indicate the presence of a thermal particle distribution or fast cooling in the inner region of the jet. In our steepest case, in Cygnus A the mid-infrared flux is strongly polarized \citep{lr14}. This supports the synchrotron nature of the spectrum, and thus we can assume that the turnover is jet-related even in the sources with the steepest spectra. While the photometric errors in the optical/IR are typically lower than $\sim$ 5\%, we considered a minimum of 10\% error on all the measured fluxes to account for variability \citep{maoz05,anderson05}. To estimate the errors on the fitted parameters we used a bootstrapping method to generate synthetic datasets. The variations on the original dataset are based on the size of the flux errors. Each one of the synthetic spectra is fitted and in the end the variance of the large number of fit results is used as the error for the fit parameters. We only take the sub-arcsecond resolution measurements into account when fitting the data, with the exception of NGC 1052. In this source the spectrum is clearly dominated by the active nucleus below $\sim$3$\times$10$^{13}$ Hz, and thus we included the low-spatial resolution data at these frequencies for a better coverage of the spectral break. The SEDs, with their best fit models, are shown in Fig. 1. The jet break frequencies, radio and jet break fluxes from the AGN sample are tabulated in Table 2.   

\begin{figure*}
\epsscale{1.17}
\plotone{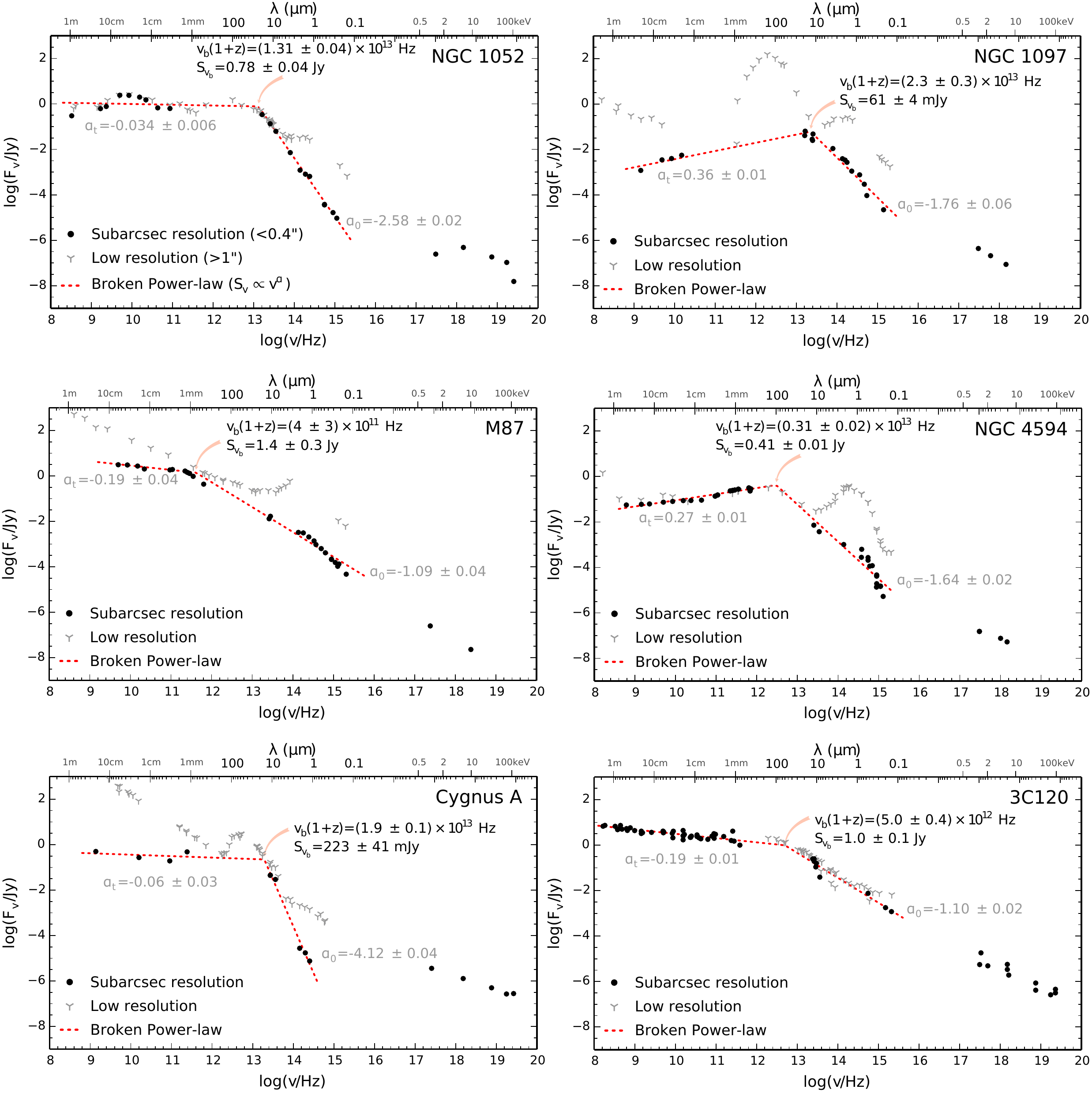}
\caption{SEDs from our sample of AGN. The low-energy part of each SED is fitted with a broken power law model (red dashed line) with four free parameters, which are the spectral slope for the optically thin part of the spectrum, $a_{0}$, the spectral slope for the optically thick part of the spectrum, $a_{t}$, the break frequency, $\nu_{b}$, and the flux at the break frequency, $S_{b}$. Low-angular resolution measurements ($>$1" apertures) are depicted as grey spikes and high-angular resolution measurements ($<$0.4'' apertures) as black dots. Only the high-angular resolution data are used to fit the model, with the exception of NGC 1052, where the low-angular resolution data are clearly dominated by the active nucleus below $\sim$3$\times$10$^{13}$ Hz.}
\end{figure*}

\subsubsection{X-ray data}

The literature values of the power law photon indices and luminosities in Eddington units typically measured from the X-ray band 2--10 keV for low-luminosity AGN \citep{gm09,terashima02} and 1--100 keV for Sgr A*, Cyg A and 3C 120 \citep{barriere14,young02,zdziarski01} of our AGN sample are tabulated in Table 2. The selected X-ray power law photon indices are less than one sigma away from the mean value as calculated from several other values found in the literature. Thus, the non-simultaneity of the X-ray observations with the radio/optical/UV does not present sizeable correction to the power law indices. Unlike for XRBs, the bolometric correction factors to turn the X-ray luminosities to accretion rates in Eddington units are larger as most of the accretion luminosity is radiated in lower wavelength regimes instead of X-rays, and can range from 10--30 (NGC 1097, NGC 4594, M87; \citealt{ho99}) to 1000 (Sgr A*; \citealt{barriere14}). 

\begin{table*}\centering
\caption{The jet break values, radio and jet break fluxes and literature values for power law photon indices and X-ray luminosities of the AGN sample. The columns are: (1) Source name, (2) AGN classification, (3) redshift, (4) logarithm of the redshift corrected jet break frequency, (5) flux density at the jet break, (6) radio flux density at 5 GHz, (7) the excess luminosity over the radio luminosity up to the jet break $L_{\rm b}/L_{\rm 5GHz} = \nu_{b} S_{\nu,b}/\nu_{\rm 5GHz} S_{\rm \nu,5GHz}$, (8) X-ray power law photon index and its 90\% error, (9) 2--10 keV unabsorbed X-ray luminosity in units of Eddington luminosity, and (10) references for the power law photon indices, X-ray luminosities and black hole masses (T02: \citet{terashima02}, W02: \citet{woo02}, L06: \citet{lewis06}, G09: \citet{gm09}, B14: \citet{barriere14}, Y02: \citet{young02}, T03: \citet{tadhunter03}, Z01: \citet{zdziarski01}, P14: \citet{pozonunez14}). The jet break values, radio and jet break fluxes of Sgr A* are taken from \citet{becket02}.} 
\label{AGNs}
\begin{tabular}{llcccccccc} \tableline
Source & Type & $z$ & log $\nu_{b}$(1+$z$) & S$_{\nu,b}$ & S$_{\rm \nu,5 GHz}$ & log $L_{\rm b}/L_{\rm 5GHz}$ & $\Gamma$ & $L_{X}/L_{\rm Edd}$ & Ref \\
Name & & & Hz & mJy & mJy \\ \tableline
NGC 1052 & LINER 1.9 & 0.005 & 13.12$\pm0.01$ & 780$\pm40$ & 2360$\pm236$ & 2.94$\pm$0.08 & 1.67$\pm0.40$ & 2$\times$10$^{-5}$ & T02, W02 \\
NGC 1097 & LINER 1 & 0.004 & 13.36$\pm0.06$ & 61$\pm4$ & 3.3$\pm0.3$ & 4.93$\pm$0.12 & 1.66$\pm0.12$ & 3$\times$10$^{-6}$ & T02, L06 \\  
M 87 & LINER 1 & 0.004 & 11.60$^{+0.25}_{-0.60}$ & 1400$\pm300$ & 3160$\pm316$ & 1.55$\pm$0.56 & 2.40$\pm0.11$ & 2$\times$10$^{-6}$ & G09 \\ 
NGC 4594 & LINER 2 & 0.003 & 12.49$\pm0.03$ & 410$\pm20$ & 74$\pm7$ & 3.53$\pm$0.09 & 1.89$\pm0.17$ & 10$^{-6}$ &T02 \\    
Sgr A* & & 0 & 11.90$\pm0.30$ & 4000$\pm1000$ & 750$\pm150$ & 2.93$\pm$0.28 & 2.35$\pm0.18$ & 7$\times$10$^{-12}$ & B14 \\
Cyg A & FRII & 0.056 & 13.28$\pm0.02$ & 223$\pm41$ & 373$\pm37$ & 3.36$\pm$0.14 & 1.52$\pm0.12$ & 0.001 & Y02, T03 \\
3C 120 & FSRQ & 0.03 & 12.71$\pm0.03$ & 1000$\pm100$ & 3620$\pm362$ & 2.45$\pm$0.12 & 1.85$\pm0.05$ & 0.009 & Z01, P14 \\
\tableline
\end{tabular}
\end{table*}

\subsection{X-ray spectral fitting}

We analyzed the X-ray data of our black hole XRB sample taken within a few days of each radio spectrum, and systematically fit each X-ray spectrum using standard phenomenological models that include a disc blackbody from an accretion disc and a power law with a high energy cut-off. The \rxtepca\/, \rxtehexte\/ and \swiftxrt\/ (when available) spectra were fitted in \isis\/ \citep{houck00} simultaneously with a suitable spectral model (see Table 3), with a constant offset between the spectra from different detectors to correct for calibration differences (if both \rxtehexte\/ clusters were present, these were added as separate data sets with individual constants). The best fit and its 90\% errors on the parameters were determined simultaneously by \isis\/ until the $\chi^{2}$ was sufficiently converged. We also estimated the X-ray luminosity in Eddington units (L$_{X}$/L$_{\rm Edd}$) by integrating the unabsorbed X-ray flux from 3.5--200 keV (normalizing the \rxtehexte\/ spectra to the level of the \rxtepca\/ spectra) with estimates of the distances and masses of the black holes \citep{russet13a}. For both MAXI J1659--152 and MAXI J1836--194 we adopt a distance of 8 kpc and the mass of the black hole as 10 M$_{\odot}$ (in \citealt{russet14} they estimated the mass of the black hole as 4--15 M$_{\odot}$, and the distance as 4--10 kpc). L$_{X}$/L$_{\rm Edd}$ is proportional to $\dot{m}$ given a bolometric correction factor that relates L$_{X}$ to bolometric luminosity of the whole accretion process. The bolometric correction factor is usually unknown, but in the case of XRBs most of the accretion luminosity is radiated in the X-ray regime, and thus the value of the bolometric correction factor is likely small (1--5). Obviously, the obtained L$_{X}$/L$_{\rm Edd}$ values are crude estimates and should be taken as accurate to an order of magnitude. The X-ray model parameters for the AGN sample were taken from the literature (see Table 2).  

\begin{table*} \centering
\caption{X-ray spectral fits of XRBs. The columns are: (1) source name, (2) the ID number of the \rxte\/ pointing, (3) the ID number of the \swift\/ pointing, (4) starting time of the pointings in MJD (\rxte/\swift), (5) the model fitted to the data, (6) the column density n$_{H}$ of the absorption component \phabs\/ in units of 10$^{22}$ cm$^{-2}$, (7) the best fit value of the X-ray power law photon index and its 90\% confidence interval, (8) the reduced $\chi^{2}$ value and the number of degrees of freedom of the fit, and (9) 3.5--200 keV unabsorbed X-ray luminosity in units of Eddington luminosity. Model key: D - \diskbb, PL - \powerlaw, CPL - \cutoffpl, G - \gaussian, E - \edge.} 
\label{models}
\begin{tabular}{lllllcccc} \tableline
Source & \rxte\/ ObsID & \swift\/ ObsID & Time (MJD) & Model & n$_{H}$ & $\Gamma$ & $\chi^{2}_{red}$/d.o.f & L$_{X}$/L$_{\rm Edd}$\\ \tableline
4U 1543-47 & 70124-02-12-00 & & 52490.12 & PL+G & 0.25$^{f}$ &1.73$\pm0.03$ & 0.70/39 & 2$\times$10$^{-3}$ \\	
Cyg X-1 & 91096-01-06-00 & & 53513.04 & EE(CPL+G) & 0.6$^{f}$ & 1.70$\pm0.01$ & 1.34/121 & 8$\times$10$^{-3}$ \\	
GS 1354-64 & 20431-01-03-00 & & 50774.39 & E(CPL+G) & 0.9$^{f}$ & 1.39$\pm0.01$ & 1.35/111 & 6$\times$10$^{-1}$ \\	
GX 339-4 & 20181-01-05-00 & & 50636.34 & E(PL+G) & 0.6$^{f}$ & 1.55$\pm0.01$ & 1.08/70 & 2$\times$10$^{-2}$ \\
& 95409-01-09-03 & & 55266.78 & E(PL+G) & 0.6$^{f}$ & 1.57$\pm0.01$ & 1.83/52 & 2$\times$10$^{-1}$ \\	
& 96409-01-09-00 & & 55617.54 & E(PL) & 0.6$^{f}$ & 1.62$\pm0.04$ & 0.78/50 & 7$\times$10$^{-3}$ \\
MAXI J1659-152 & 95358-01-02-00* & 00434928005 & 55467.04/55467.30 & D+PL+3G & 0.30$\pm0.01$ & 1.93$\pm0.01$ & 1.63/403 & 7$\times$10$^{-2}$ \\
& 95358-01-02-01 & 00434928007 & 55468.08/55468.22 & D+PL+3G & 0.33$\pm0.01$ & 2.08$\pm0.01$ & 1.42/237 & 5$\times$10$^{-2}$ \\		
& 95358-01-03-00 & 00434928009 & 55470.24/55470.24 & D+PL+3G & 0.32$\pm0.01$ & 2.17$\pm0.01$ & 1.49/292 & 6$\times$10$^{-2}$ \\
& 95108-01-05-00 & 00434928011 & 55472.07/55472.11 & D+PL+3G & 0.34$\pm0.01$ & 2.20$\pm0.03$ & 1.05/203 & 5$\times$10$^{-2}$ \\
& 95108-01-11-00* & 00434928013 & 55474.57/55474.12 & D+PL+3G & 0.32$\pm0.01$ & 2.24$\pm0.03$ & 1.32/295 & 5$\times$10$^{-2}$ \\
& 95108-01-18-01 & 00434928017 & 55477.00/55477.12 & D+PL+G & 0.39$\pm0.01$ & 2.15$\pm0.02$ & 1.59/237 & 6$\times$10$^{-2}$ \\
& 95118-01-06-00* & 00031843003 & 55489.26/55489.04 & D+PL & 0.33$\pm0.01$ & 2.15$\pm0.03$ & 1.65/269 & 2$\times$10$^{-2}$ \\
MAXI J1836-194 & 96371-03-03-00* & 00032087002 & 55806.48/55805.23 & D+PL+G & 0.32$\pm0.03$ & 2.13$\pm0.02$ & 1.15/137 & 1$\times$10$^{-2}$ \\
& 96438-01-02-00* & 00032087013 & 55821.84/55821.69 & D+PL+G & 0.33$\pm0.01$ & 2.38$\pm0.06$ & 1.53/189 & 1$\times$10$^{-2}$ \\	
& 96438-01-03-05 & 00032087017 & 55830.88/55830.18 & D+PL+G & 0.23$\pm0.03$ & 1.82$\pm0.02$ & 1.28/125 & 2$\times$10$^{-2}$ \\	
& 96438-01-05-05 & 00032087024 & 55846.55/55846.84 & D+PL+G & 0.16$\pm0.06$ & 1.56$\pm0.02$ & 1.40/70 & 9$\times$10$^{-3}$ \\	
& 96438-01-07-04 & 00032087029 & 55861.51/55861.21 & PL+G & 0.19$\pm0.02$ & 1.63$\pm$0.02 & 0.79/55 & 9$\times$10$^{-3}$ \\	
V4641 Sgr & 80054-08-01-01*$\dag$ & & 52857.37 & EE(PL+G) & 0.4$^{f}$ & 0.93$\pm0.03$ & 1.10/28 & 3$\times$10$^{-2}$\\	
XTE J1118+480 & 50137-01-06-00 & & 51649.04 & PL+G & 0.01$^{f}$ & 1.72$\pm0.01$ & 1.10/74 & 1$\times$10$^{-3}$ \\
& 90011-01-01-08 & & 53386.34 & PL & 0.01$^{f}$ & 1.76$\pm0.01$ & 0.95/56 & 5$\times$10$^{-4}$ \\
XTE J1550-564 & 50135-01-12-00 & & 51696.47 & E(PL+G) & 0.65$^{f}$ & 1.59$\pm0.01$ & 0.72/55 & 4$\times$10$^{-3}$ \\	
XTE J1752-223 & 95702-01-11-01* & & 55377.21 & PL & 0.65$^{f}$ & 1.87$\pm0.06$ & 1.12/29 & 6$\times$10$^{-3}$ \\
\tableline
\end{tabular}
\begin{list}{}{}
\item[*] Only \rxtepca\/ spectrum is used for the fit. \\
\item[$\dag$] The first part of this pointing exhibits strong flares and rapid spectral variability \citep{maitra06}, and thus we select the second part of the pointing for fitting which is more stable. \\
\item[$^{f}$] Fixed value in the model fitting. \\
\end{list}
\end{table*}

\section{Results}

Comparing the jet properties with the properties of the plasma close to the black hole, we found a relationship among the X-ray power law photon index, the jet break frequency and the ``excess luminosity'' over the radio luminosity (see below). The photon indices for black hole systems are plotted against the break frequencies in Fig. 2 (circles and vertical bars representing XRBs with one individual XRB, MAXI J1836--194, highlighted as orange, and green triangles representing AGN). There is a clear anti-correlation in the data, encompassing the sample of XRBs, an individual XRB, and the sample of AGN, spanning five orders of magnitude in the jet break frequency. To better quantify this anti-correlation we calculated the correlation coefficient using Monte Carlo methods \citep{curran14} (Table 4). The Monte Carlo method involves creating M (here M=10$^{6}$) new data sets based on the original data that are composed of N data pairs of $\Gamma$ and $\nu_{b}$. Each new data set consists of randomly chosen pairs from the original data, such that some of the original pairs may appear more than once or not at all. In addition, the pairs in a given new data set are randomly perturbed by random sampling from the normal distribution with means and standard deviations according to the original pairs, or from the uniform distribution in the case of a range of frequencies as described above. Taking into account the whole data set, the correlation coefficient amounted to $-0.75$ with a significance of 4.6$\sigma$ (see Table 4 for statistics of subsets). We also performed a linear least-squares regression which is calculated for all randomly perturbed samples and we note the slope and intercept of each fit, and produce 95\%, 99\% and 99.9\% limits on the possible regressions: $\log~(\nu_{b}/\mathrm{Hz}) = -3.4^{+0.9}_{-1.4} \Gamma + 18.8^{+2.5}_{-1.6}$ (the errors are 95\% confidence interval; see also Fig. 2). This anti-correlation clearly shows an intimate connection between the region near the black hole (responsible for the X-ray emission) and the jet emission (responsible for the $\nu_{\rm break}$). 

\begin{figure*}
  \epsscale{1.18}
  \plotone{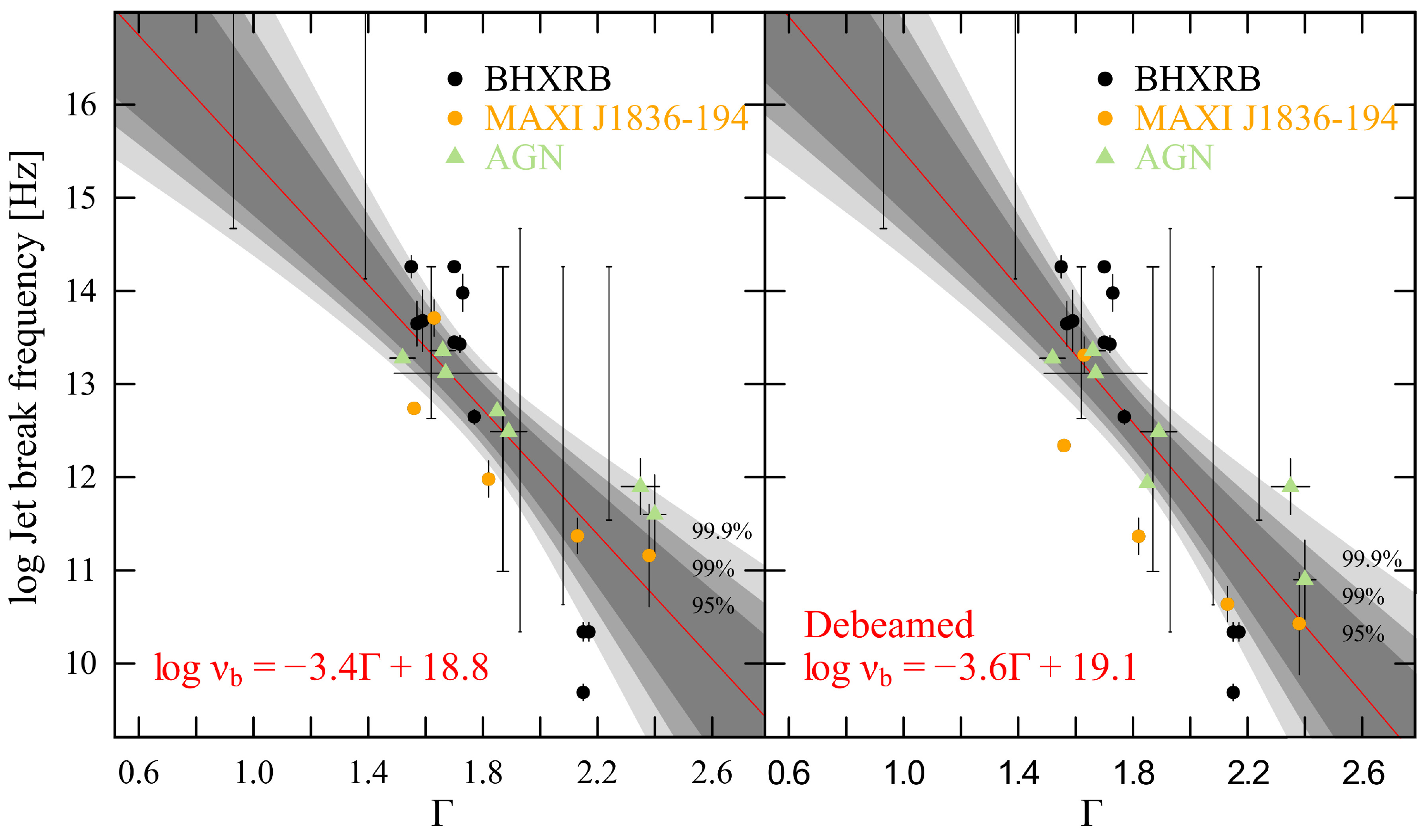}
\caption{\textit{Left:} The jet break frequency as a function of the X-ray power law photon index for black hole systems. \textit{Right:} The same as in the left panel but the jet breaks have been corrected for beaming. In both panels black circles and vertical bars represent the data from XRBs and green triangles from AGN. One individual XRB MAXI J1836--194 is highlighted as orange with multiple observations across a state change from hard to intermediate X-ray state. The red solid line shows the median of the Monte-Carlo bootstrap samples with its formula written down in the lower left corner of the figure. The shaded grey regions show the 95\%, 99\% and 99.9\% confidence intervals on the linear regressions. The XRB data with only upper or lower limits on the jet break frequency available are shown as vertical bars depicting the range of the frequency we are considering as a conservative estimate of the break frequency, with the horizontal bars at the ends of the vertical bar depicting the 1$\sigma$ error on the X-ray power law photon index.}
\end{figure*}

\begin{table}\centering
\caption{Spearman rank coefficients for the break frequency vs. X-ray power law photon index correlation. We use ten different samples to study the correlation: XRBs, AGN and both of them combined (denoted as ``ALL'') including the observations with ranges on the break frequency, the samples that do not include the ranges (marked with a star [*]), and the samples that are debeamed (marked with letters `db'). The correlation coefficient and the best linear fit and their confidence limits are determined by Spearman's rank Monte-Carlo bootstrapping, where 10$^{6}$ sets of simulated data are created by the bootstrap function and we note the most likely correlation coefficient, 95\% confidence interval on the correlation coefficient, and the percentage of cases where the null hypothesis is valid (see Methods for more detailed discussion). The columns are: (1) sample name, (2) the Spearman rank correlation coefficient, (3) 95\% confidence limit on the correlation coefficient, (4) the percentage of cases where the null hypothesis is valid for Monte Carlo bootstrap method.} 
\label{correlations}
\begin{tabular}{rcccccc}\tableline
& R & 95\% conf. & null (\%) \\
\tableline
ALL & -0.76 & (-0.87)--(-0.52) & 2$\times$10$^{-4}$ \\
ALL* & -0.76 & (-0.88)--(-0.48)& 4$\times$10$^{-3}$ \\ 
ALL(db) & -0.75 & (-0.86)--(-0.50) & 8$\times$10$^{-4}$ \\
ALL(db)* & -0.75 & (-0.88)--(-0.47) & 6$\times$10$^{-4}$ \\
XRB & -0.77 & (-0.89)--(-0.46) & 8$\times$10$^{-3}$ \\
XRB* & -0.76 & (-0.91)--(-0.39) & 0.08 \\
XRB(db) & -0.76 & (-0.89)--(-0.44) & 0.01 \\
XRB(db)* & -0.76 & (-0.91)--(-0.37) & 0.1 \\
AGN & -0.86 & (-1.00)--(-0.32) & 0.6 \\
AGN(db) & -0.86 & (-0.96)--(-0.29) & 0.7 \\
\tableline
\end{tabular}
\end{table}

With the knowledge of the frequency and the flux density of the jet break, we can estimate the excess luminosity caused by the variable break frequency over the radio luminosity by measuring $L_{\rm b}/L_{\rm 5GHz} = \nu_{b} S_{\nu,b}/\nu_{5GHz} S_{\nu,5GHz} = (\nu_{b}/\nu_{5GHz})^{1+\alpha_{thick}}$, which varies between sources by six orders of magnitude (see Tables 1 and 2). This result clearly demonstrates that the jet luminosities should be recalibrated taking the break frequency into account. Due to the relation between the break frequency and the X-ray power law photon index, the excess luminosity depends on the value of the X-ray power law photon index, and can be estimated as $\log~(L_{\rm b}$/erg\,s$^{-1})$ = $\log~L_{\rm 5GHz} - 3.5^{+0.9}_{-1.0} \Gamma + 9.8^{+2.0}_{-1.6}$ (the errors are 95\% confidence interval) by performing a linear least-squares regression using the above Monte Carlo methods between the excess luminosity and X-ray power law photon index.

\subsection{The effect of beaming}

Generally, it is thought that the bulk Lorentz factors of the jets in XRBs should be around two \citep{fender04,casella10}. However, direct observational evidence is largely missing in this regard. Together with the inclinations that are generally $>$30$^{\circ}$ the Doppler factors of our XRB sample can be estimated as $\delta\sim$ 1, apart from one source. MAXI 1836--194 is suspected to have a slightly higher bulk Lorentz factor ($\Gamma_{b}$ = 1--4) and a jet angle oriented close to our line-of-sight (4$^{\circ}$--15$^{\circ}$, \citealt{russet15}). It is likely that the Doppler factor is variable depending on the X-ray flux of the source (see Fig. 10 in \citealt{russet15}). We use their estimation of the bulk Lorentz factors depending on the X-ray flux and the jet angle of 10$^{\circ}$, and correct for the effect of Doppler boosting ($\delta=$ 2--5) in Fig. 2 (right panel) and Table 4. 

As in XRBs, we consider the effect of beaming to be small for our sample of AGN. The only beaming candidates in our sample are M87 and 3C 120, where the Doppler factors have been estimated to be 2--5 \citep{wanget09} and 5.9 \citep{hovaet09}, respectively. Similar to the case of MAXI 1836--194 mentioned above, we correct for the effect of Doppler boosting for M87 and 3C 120 in Fig. 2 (right panel) and Table 4. In general, the debeaming does not have a big impact on the correlation.

\section{Discussion}

In the above we have shown that there exists an intimate connection between the jet break frequency and X-ray power law photon index. The observed optically thick radio spectra can be produced by many models ranging from non-thermal, hybrid, or thermal distributions of electrons in a single acceleration episode, distributed acceleration along the jet, or an internal shock model \citep{blanko79,peer09,falcet00,stawet02,bottcet10,malzac13}. Thus, more observables are clearly needed to single out a favorable scenario, and observations around and above the jet break are crucial in this regard. With the observed correlation we can state that the conditions dictating the jet break and jet spectrum are set by the X-ray emitting region, or that they are both driven by an underlying parameter. This idea is further supported by the correlation found earlier between the radio and X-ray luminosities \citep{hannet98,corbet03,gallet03,fendga14}. The hard X-ray emission is most likely produced by inverse Compton scattering, although a synchrotron origin has been suggested for some individual XRBs in the hard X-ray state for very low accretion rates ($\dot{m} \sim 10^{-4}$; \citealt{russet10}); even in this case, the X-ray power law photon index is similar to that produced by inverse Compton scattering. In the future, broadband observations of quiescent, low accretion rate black hole XRBs would be an ideal test of the validity and extent of the correlation. The inverse Compton spectrum depends on the energy distribution of the electrons, optical depth of the medium and the energy density of the seed photons. Qualitatively, the correlation could be explained by an increase in the amount of seed photons that would produce more Compton scattering. The increased scattering cools down the electron population close to the black hole that is then translated to decreasing values of the jet break, in similar fashion as the blazar sequence \citep{ghiset98}.

Different semi-analytical models and magnetohydrodynamical simulations of jet formation assume different initial conditions, e.g. magnetically arrested disk (MAD; \citealt{tcheet11}) simulations assume that jet launching is magnetically dominated whereas RIAF simulations assume it is not \citep{yuanet03}. How magnetically dominated the jets are, what they contain, and how particles are accelerated, are all uncertain at this stage. Thus, the initial conditions for jet formation are currently something that all forms of modeling, semi-analytical as well as simulations (e.g. particle-in-cell simulations; \citealt{sironi14}), essentially have to insert by hand. Our result provides an observational connection between the characteristics of the particle distribution and properties of the jet, and thus may narrow down the initial condition parameter space for jet formation models.

\subsection{The effect of the black hole mass}

Due to the mass-scaling properties of black holes, it has been suggested that accretion physics scales globally with black hole mass \citep{heinz03}. Such a scaling is supported by the discovery of the Fundamental Plane of black hole activity, an observed relationship between the X-ray luminosity, radio luminosity and black hole mass in the hard X-ray state (i.e., compact jet producing) black hole XRBs and AGN \citep{merlet03,falcet04,plotet12}. According to theoretical scaling relations \citep{heinz03}, the break frequency between the flat/inverted power law and optically thin power law scales with the dimensionless accretion rate $\dot{m}$ (defined as the mass accretion rate divided by the Eddington rate) and black hole mass $M_{BH}$ as $\nu_{b} \propto M_{BH}^{-(p+2)/(2p+8)} \dot{m}^{(p+6)/(2p+8)}$ for sources with $\dot{m}\lesssim$ a few percent, which reduces to $\nu_{b} \propto M_{BH}^{-1/3} \dot{m}^{2/3}$ assuming the electron distribution is a power law with index $p=2$ (the above scaling relation is not very sensitive to the value of p). In this case, the difference between AGN and XRBs in $\nu_{b}$ is predicted to be three orders of magnitude in frequency (for nominal values of $M_{BH,AGN} \sim $10$^{9} M_{\odot}$ and $M_{BH,XRB} \sim 10M_{\odot}$) for the same $\dot{m}$. The difference would be even larger if we take the mass accretion rate into account and assume that the XRBs have systematically larger $\dot{m}$ than AGN, which is likely the case as the XRBs in our sample are fairly luminous, with $\dot{m}>10^{-3}$. Such a large difference is not observed for black hole masses in the range from 10$M_{\odot}$ to 6$\times$10$^{9}M_{\odot}$ covered by our sample, so alternative explanations are required. 

\subsection{The effect of the mass accretion rate}

It is known that the mass accretion rate does not vary substantially over the state transition for XRBs -- the X-ray flux, which is a proxy for the mass accretion rate, stays at a similar level when XRBs make a transition from the hard X-ray state to the soft X-ray state and vice versa \citep{kordet06}. However, according to the correlation between the jet break and X-ray power law photon index, the largest change in both parameters occur specifically during the transition with $\Gamma$ changing from $\sim1.6$ to $\sim2.4$ and $\nu_{b}$ from $\sim10^{14}$ Hz to $\sim10^{11}$ Hz, whereas during the hard state ($\Gamma \sim 1.6$) the break frequency and X-ray power law photon index remain approximately constant while the mass accretion rate is changing by orders of magnitude. In addition, one XRB in the sample (MAXI J1836--194) shows a change in the frequency of the jet break by three orders of magnitude over state transitions (orange dots in Fig. 2), demonstrating that the effect of the black hole mass and spin on the jet properties including the jet power is negligible compared to the accretion state changes. Therefore the observed correlations could provide evidence for the internal physics and/or accretion mode being the jet power driver for both XRBs and low accretion rate AGN. 

\subsection{Estimating the total jet luminosity}

Estimating the total jet luminosity (ignoring beaming) from the jet spectral energy distribution requires also the knowledge of the flux above the jet break frequency. Above the break the optically thin spectrum reveals the underlying particle distribution: either a power law ($\alpha_{thin} \sim -0.7$ or steeper depending on the cooling mechanisms) or a quasi-thermal distribution ($\alpha_{thin} < -1.0$). If the optically thin spectrum is shallow enough ($\alpha_{thin} > -1.0$) the jet luminosity then depends on the emission above the break frequency, up to the electron cooling break frequency. In some black hole XRBs the optically thin spectrum is characterized with $\alpha_{thin} > -1.0$, however, in only one case is there reliable evidence of how far the optically thin spectrum extends \citep{russet14}, corresponding roughly to $10^{3}\nu_{b}$. As a conservative estimate, the total luminosity normalized to the 5 GHz radio luminosity for XRBs would be less than one order of magnitude greater than $L_{\rm b}/L_{\rm 5GHz}$ when taking the mean optically thin spectral index ($\alpha_{thin} \sim -0.85$) from our sample of XRBs when constrained, and assuming that the cooling break is located at $\leq 10^{4}\nu_{b}$. Thus, we can consider that the excess luminosity gives an order of magnitude estimate of the total jet luminosity.
 
\section{Conclusions}

We have collected an unprecedented data set of multiwavelength spectral energy distributions from the core of the compact jet in stellar-mass and supermassive black holes, in addition to (near-simultaneous in the case of XRBs) X-ray observations. We have discovered a correlation between the X-ray power law photon index of the corona and the jet break frequency, and a resulting correlation between the X-ray power law photon index and the ``excess luminosity'' over the radio luminosity, suggesting an intrinsic connection between the plasma close to the black hole and the outflow properties. Further considerations are needed to determine the nature of the jet break and the spectral slope of the optically thin part of the jet spectrum, which can be achieved by detailed modeling of the spectral energy distributions of the sources. Our results indicate that the jet production and properties (possibly coupled to the X-ray spectral state changes) are closely related to changes in the corona/hot flow. Therefore this result will serve as a benchmark that should be reproduced by the jet formation models, and provides observational clues about the connection between particle acceleration and properties of the jet. Our results support the notion that the coronae are black hole mass- and spin-independent features of black hole accretion whose presence are essential in producing powerful jets on all scales. 

\acknowledgments

This research has made use of data obtained from the High Energy Astrophysics Science Archive Research center (HEASARC), provided by NASA's Goddard Space Flight center. SM is grateful to the University of Texas in Austin for its support through a Tinsley Centennial Visiting Professorship. PAC acknowledges support from Australian Research Council grant DP120102393.






\bibliography{refs1}

\clearpage

\end{document}